\documentclass[12pt]{article}

\usepackage{amsmath,amssymb}
\usepackage{graphicx}

\makeatletter

\renewcommand\section{\@startsection{section}{1}{\z@}
                                   {-3.5ex \@plus -1ex \@minus -.2ex}
                                   {2.3ex \@plus .2ex}
                                   {\normalfont\large\bfseries}}
\renewcommand\subsection{\@startsection{subsection}{2}{\z@}
                                   {-3.25ex\@plus -1ex \@minus -.2ex}
                                   {1.5ex \@plus .2ex}
                                   {\normalfont\normalsize\bfseries}}
\renewcommand\subsubsection{\@startsection{subsubsection}{3}{\z@}
                                   {-3.25ex\@plus -1ex \@minus -.2ex}
                                   {1.5ex \@plus .2ex}
                                   {\normalfont\normalsize\bfseries}}
\renewcommand\paragraph{\@startsection{paragraph}{4}{\z@}
                                   {3.25ex \@plus1ex \@minus.2ex}
                                   {-1em}
                                   {\normalfont\normalsize\bfseries}}

\makeatother

\newcommand{\beq}{\begin{equation}}
\newcommand{\eeq}{\end{equation}}
\newcommand{\bea}{\begin{eqnarray}}
\newcommand{\eea}{\end{eqnarray}}

\newcommand{\SL}{{\rm SL}}
\newcommand{\SU}{{\rm SU}}
\newcommand{\SO}{{\rm SO}}
\newcommand{\Sp}{{\rm Sp}}

\newcommand{\R}{\mathbb R}
\newcommand{\Z}{\mathbb Z}

\newcommand{\id}{\hbox{1\kern-.27em l}}
\newcommand{\Gadj}{G_{\rm adj}}
\newcommand{\gadj}{g_{\rm adj}}

\newcommand{\Tr}{{\rm Tr}}

\newcommand{\Ad}{{\rm Ad}}
\newcommand{\ad}{{\rm ad}}

\newcommand{\tp}{{\tilde{p}}}
\newcommand{\tx}{{\tilde{x}}}

\newcommand{\cA}{{\cal A}}

\newcommand{\cD}{{\cal D}}

\newcommand{\cH}{{\cal H}}

\newcommand{\cL}{{\cal L}}
\newcommand{\cM}{{\cal M}}

\newcommand{\cO}{{\cal O}}

\begin{document}

\pagestyle{empty}

\begin{center}

\vspace*{30mm}
{\Large  BPS states in $(2, 0)$ theory on $\R \times T^5$}

\vspace*{30mm}
{\large M{\aa}ns Henningson}

\vspace*{5mm}
Department of Fundamental Physics\\
Chalmers University of Technology\\
S-412 96 G\"oteborg, Sweden\\[3mm]
{\tt mans@chalmers.se}     
     
\vspace*{30mm}{\bf Abstract:} 
\end{center}
We consider $(2, 0)$ theory on a space-time of the form $\R \times T^5$, where the first factor denotes time, and the second factor is a flat spatial five-torus. In addition to their energy, quantum states are characterized by their spatial momentum, 't~Hooft flux, and $\Sp (4)$ $R$-symmetry representation. The momentum obeys a shifted quantization law determined by the 't~Hooft flux. By supersymmetry, the energy is bounded from below by the magnitude of the momentum. This bound is saturated by BPS states, that are annihilated by half of the supercharges. The spectrum of such states is invariant under smooth deformations of the theory, and can thus be studied by exploiting the interpretation of $(2, 0)$ theory as an ultra-violet completion of maximally supersymmetric Yang-Mills theory on $\R \times T^4$. Our main example is the $A$-series of $(2,0)$ theories, where such methods allow us to study the spectrum of BPS states for many values of the momentum and the 't~Hooft flux. In particular, we can describe the $R$-symmetry transformation properties of these states by determining the image of their $\Sp (4)$ representation in a certain quotient of the $\Sp (4)$ representation ring.

\newpage

\pagestyle{plain}

\setcounter{equation}{0}
\section{Introduction}
Understanding the conceptual foundations of the six-dimensional quantum theories with $(2, 0)$ supersymmetry \cite{Witten95} remains, at least for the present author, an outstanding challenge. In this paper, we will consider such a theory defined by an element $\Phi$ of the $ADE$-classification on a space-time of the form
\beq
\R \times T^5 = \R \times \R^5 / \Lambda .
\eeq
Here the factor $\R$ denotes time, we identify the spatial $\R^5$ factor with its dual $(\R^5)^*$ by means of the standard flat metric, and $\Lambda \subset \R^5$ is a rank five lattice. 

A basic question is what the possible values of the spatial momentum $p$ are. One might think that these should be given by the lattice
\beq
\Lambda^* \simeq H^1 (T^5, \Z) \subset H^1 (T^5, \R)
\eeq
dual to $\Lambda$, but this is not quite true. To explain this point, we consider a further discrete quantum number
\beq
f \in H^3 (T^5, C)
\eeq
known as the 't~Hooft flux. Here the finite abelian group $C$ is isomorphic to the center subgroup of the simply connected Lie group $G$ corresponding to the element $\Phi$ of the $ADE$-classification. Thus
\beq
C \simeq \Gamma_{\rm weight} / \Gamma_{\rm root} ,
\eeq
where $\Gamma_{\rm weight}$ and its dual $\Gamma_{\rm root}$ are the weight- and root-lattices of $G$ respectively. The inner product on the weight space of $G$ induces a perfect pairing on $C$ with values in $\R / \Z \simeq U (1)$. As we will explain in more detail later, it is then possible to define a product
\beq
f \cdot f \in H^1 (T^5, \R / \Z) ,
\eeq
and the correct quantization law for the momentum $p \in H^1 (T^5, \R)$ can be shown to be
\beq \label{p-quantization}
p - f \cdot f \in H^1 (T^5, \Z) \simeq \Lambda^* .
\eeq

The main theme of this paper is to analyze the implications of supersymmetry. The generators $Q^i$, $i = 1, \ldots, 4$ of infinitesimal supersymmetries transform in the fundamental representation ${\bf 4}$ of the $\Sp (4)$ $R$-symmetry group. In six-dimensional Minkowski space, they also transform as a Weyl spinor under the $\SO (1, 5)$ Lorentz group, and obey a symplectic Majorana condition. (On $\R \times T^5$, the Lorentz group is of course broken to a discrete (and generically trivial) subgroup of the $\SO (5)$ spatial rotation group, so actually only the $\Sp (4)$ representation content ${\bf 4} \oplus {\bf 4} \oplus {\bf 4} \oplus {\bf 4}$ of the supercharges is relevant. But to avoid cluttering the notation, it is still convenient to present some formulas in an $\SO (1,5)$ covariant way.) The equal-time anti-commutation relations of the supercharges can be written in the form
\beq \label{susy-algebra}
\{ Q^i, Q_j^\dagger  \} = \delta^i_j \left( E \id - \gamma^0 \gamma \cdot p \right) ,
\eeq
where $E$ and $p$ denote the energy and the momentum respectively. ($\gamma^0$ and $\gamma$ are the temporal and spatial Dirac matrices). Unitarity requires the matrix on the right hand side to be positive semi-definite, from which follows that
\beq
E \geq \left| p \right| .
\eeq
We may thus distinguish between three broad classes of states: 
\begin{itemize}
\item
{\it Vacuum states} have 
\beq
E = p = 0 ,
\eeq
so that the right hand side of (\ref{susy-algebra}) is identically zero. In view of (\ref{p-quantization}), a necessary requirement for such a state is that $f \cdot f = 0$. It is then consistent to impose that these states are annihilated by all supercharges. The spectrum of such states  was investigated in \cite{Henningson} (in greatest detail for the $A$- and $D$-series).
\item 
{\it BPS states} generically have 
\beq
E = \left| p \right| > 0 , 
\eeq
so that the right hand side of (\ref{susy-algebra}) has half maximal rank. Indeed, 
\beq
E \id - \gamma^0 \gamma \cdot p = E \left( \id - \gamma_p \right) ,
\eeq
where the transverse chirality matrix $\gamma_p$ defined by
\beq
\gamma_p = \left| p \right|^{-1} \gamma^0 \gamma \cdot p
\eeq
has eigenvalues $+1$ and $-1$ with equal multiplicities. It is then consistent to impose that these states are annihilated by the supercharges with positive chirality. To understand the structure of such a multiplet, it is convenient to regard the remaining supercharges of negative chirality as a set of fermionic creation operators transforming in the ${\bf 4}$ representation of $\Sp (4)$ and the corresponding annihilation operators (also in the ${\bf 4}$ representation). For given values of the momentum $p$ and the 't~Hooft flux $f$, we start with a set of states which are annihilated by the annihilation operators and transform in some (in general reducible) representation $R_{(f,p)}$ of $\Sp (4)$. Acting with the creation operators then builds up a multiplet of states transforming as
\beq \label{BPS-multiplet}
(B \oplus F) \otimes R_{(f,p)} ,
\eeq
where the $\Sp (4)$ representations $B$ and $F$ (for `Bosonic' and `Fermionic'), are given as direct sums of the even and odd alternating powers of ${\bf 4}$ respectively:
\bea \label{BF}
B & = & {\bf 1} \oplus ({\bf 4})^2_a \oplus ({\bf 4})^4_a \cr
& = & {\bf 1} \oplus {\bf 1} \oplus {\bf 1} \oplus {\bf 5} \cr
F & = & {\bf 4} \oplus ({\bf 4})^3_a \cr
& = & {\bf 4} \oplus {\bf 4} .
\eea    
As usual, the spectrum of such states, as described by the $\Sp (4)$ representation $R_{(f, p)}$, can be expected to be invariant under a large class of continuous deformations of the theory, notably including deformations of the flat metric on $T^5$. There is thus some hope of determining it explicitly, at least in some cases. This is the goal of the present paper. An important point is that the representation $R_{(f,p)}$ can only depend on the orbit  $[f, p]$ of the pair $(f, p)$ under the $\SL_5 (\Z)$ mapping class group of $T^5$.  
\item
{\it Non-BPS states} have 
\beq
E > \left| p \right| ,
\eeq
 so that the right hand side of (\ref{susy-algebra}) has maximal rank. The fermionic creation operators, as well as the annihilation operators, then transform in the ${\bf 4} \oplus {\bf 4}$ representation of $\Sp (4)$, and build up multiplets of the form
\beq \label{non-BPS-multiplet}
(B \oplus F) \otimes (B \oplus F) \otimes R^\prime_{(f,p)}
\eeq
for some $\Sp (4)$ representation $R^\prime_{(f,p)}$. Understanding the structure of this representation and the corresponding energies $E$ appears out of reach at the present, though.
\end{itemize}

We will now briefly describe the strategy to compute the BPS spectrum and outline the rest of the paper. As we discuss in section two, the key point is to consider $T^5$ as a product
\beq
T^5 = T^4 \times S^1 .
\eeq
Type $\Phi$ $(2, 0)$ theory on $\R \times T^5$ can then be regarded as the ultraviolet completion of maximally supersymmetric Yang-Mills theory on $\R \times T^4$ with gauge group
\beq
\Gadj = G / C
\eeq 
of adjoint type, and a coupling constant determined by the radius of the $S^1$ factor \cite{Witten02}. The component of the momentum $p$ of $(2, 0)$ theory along the $S^1$ direction is given by the instanton number $k$ over $T^4$ of the Yang-Mills theory. The quantization law (\ref{p-quantization}) then amounts to certain topological facts concerning principal $\Gadj$ bundles over $T^4$. 

In section three, we take the weak coupling (small radius) limit. For non-zero instanton number $k$, the theory then formally reduces to a version of supersymmetric quantum mechanics on the corresponding moduli space of (anti) instanton configurations. But in practice, this model is difficult to analyze.

In section four, we instead turn to the case of zero instanton number $k = 0$. In the weak coupling limit, the theory then localizes on the moduli space of flat connections. Isolated flat connections are particularly useful, since fluctuations around them can be reliably analyzed by semi-classical methods. In this way, we can determine the contribution of such a connection to the BPS representation (\ref{BPS-multiplet}) modulo representations of the form (\ref{non-BPS-multiplet}). In other words, we may determine the image $[R_{(f,p)}]$of $R_{(f,p)}$ in the corresponding quotient of the $\Sp (4)$ representation ring. Furthermore, assuming that all representations of the form (\ref{non-BPS-multiplet}) in fact belong to non-BPS states allows us to make a concrete proposal for the actual representation $R_{(f,p)}$ of the BPS states.

I would think that it should eventually be possible to understand also the contributions from flat connections that are not isolated, but we will not pursue this here. Instead, in section five, we restrict our attention to the $A$-series, so that $G = \SU (n)$ for some $n$. The reason is that, for certain principal $\Gadj$ bundles, all flat connections are then isolated and can be treated as described above. (The case when $n$ is prime is particularly convenient.) In this way, we arrive at a proposal for the BPS representations $R_{(f, p)}$ for many values of the pair $(f, p)$ subject to (\ref{p-quantization}). 

\setcounter{equation}{0}
\section{From $(2, 0)$ theory to supersymmetric Yang-Mills theory}
\subsection{The momentum and the 't~Hooft flux}
As discussed in the introduction, the flat five-torus $T^5$ can be constructed as
\beq
T^5 = \R^5 / \Lambda ,
\eeq
where $\R^5$ is endowed with the standard flat metric, and  $\Lambda \subset \R^5$ is a rank five lattice. It follows that
\beq
\Lambda \simeq H_1 (T^5, \Z) .
\eeq

Let now $\lambda_5 \in \Lambda$ be a primitive lattice vector, which we complete to a basis $\lambda_1, \ldots, \lambda_5$ of $\Lambda$. The dual basis of the dual lattice
\beq
\Lambda^* \simeq H^1 (T^5, \Z)
\eeq
is denoted $\lambda^1, \ldots, \lambda^5$. We decompose the lattice $\Lambda$ as
\beq \label{Lambda-decomposition}
\Lambda = \tilde{\Lambda} \oplus (\lambda_5 \otimes \Z) ,
\eeq
where $\tilde{\Lambda}$ is the rank four lattice generated by $\lambda_1, \ldots, \lambda_4$. (In general, we denote four-dimensional quantities with a tilde.) After an $\SO (5)$ spatial rotation, we may assume that $\R^4 = \tilde{\Lambda} \otimes \R$ is the standard four-dimensional subspace
\beq
\R^4 = \{ (x^1, \ldots, x^5) \in \R^5 | x^5 = 0 \} 
\eeq
of $\R^5$. Finally we define a flat four-torus $T^4$ as
\beq
T^4 = \R^4 / \tilde{\Lambda} .
\eeq
So topologically, 
\beq
T^5 = T^4 \times S^1 , 
\eeq
where 
\beq
S^1 = (\lambda_5 \otimes \R) / (\lambda_5 \otimes \Z)
\eeq
is a circle in the direction of $\lambda_5$. The $(2, 0)$ theory can now be regarded as an ultra-violet completion of maximally supersymmetric Yang-Mills theory on 
\beq
\R \times T^4 ,
\eeq
where the first factor denotes time. 

The gauge group of the Yang-Mills theory is
\beq
\Gadj = G / C ,
\eeq
where the $G$ is the simply connected (and simply laced) Lie group corresponding to the element $\Phi$ of the $ADE$-classification. It is important to note that the gauge group is not simply connected (unless $\Phi = E_8$). Indeed,
\beq
\pi_1 (\Gadj) \simeq C .
\eeq
The next step in defining the theory is to choose a gauge bundle $P$, i.e. a principal $\Gadj$ bundle over $T^4$. The isomorphism class of $P$ is determined by two characteristic classes: The magnetic 't~Hooft flux
\beq
m = m_{12} \lambda^1 \cup \lambda^2 + \ldots + m_{34} \lambda^3 \cup \lambda^4 \in H^2 (T^4, C) ,
\eeq
which is the obstruction against lifting $P$ to a principal $G$-bundle over $T^4$, and the (fractional) instanton number
\beq
k  \in H^0 (T^4, \R) \simeq \R .
\eeq
The classes $m$ and $k$ may not be chosen independently, but are correlated as
\beq \label{fractional_instanton_number}
k - m \cdot m \in \Z .
\eeq
Here we have defined the product $m \cdot m \in \R / \Z$ by
\beq \label{mdotm}
m \cdot m = m_{12} m_{34} + m_{13} m_{42} + m_{14} m_{23}  ,
\eeq
with the multiplications given by the pairing on $C$. (The components of $m$ are antisymmetric in the sense that $m_{12} = - m_{21}$, et cetera in additive notation.) Clearly the definition of $m \cdot m$ is invariant under the $\SL_4 (\Z)$ mapping class group of $T^4$. (We caution the reader that the quantity $m \cdot m$ is denoted as  $\frac{1}{2} m \cdot m$ in many papers including \cite{Henningson}.)

The group $\tilde{\Omega} = {\rm Aut} (P)$ of gauge transformations may be identified with the space of sections of the associated bundle
\beq
\Ad (P) = P \times_{\Ad} \Gadj ,
\eeq
where $\Ad$ denotes the adjoint action of $\Gadj$ on itself. We let $\Omega_0$ denote the connected component of  $\tilde{\Omega}$, and define the quotient group $\Omega$ of `large' gauge transformations as
\beq
\Omega = \tilde{\Omega} / \Omega_0 \simeq {\rm Hom} (\pi_1 (T^4), C) \simeq H^1 (T^4, C) .
\eeq
A physical state must be invariant under $\Omega_0$, but may transform with non-trivial phases under $\Omega$. These transformation properties are described by the electric 't~Hooft flux
\beq
e = e_{123} \lambda^1 \cup \lambda^2 \cup \lambda^3 + \ldots + e_{234} \lambda^2 \cup \lambda^3 \cup \lambda^4 \in H^3 (T^4, C^*)  
\eeq
of the state. Here 
\beq
C^* = {\rm Hom} (C, U(1))
\eeq
is the Pontryagin dual of $C$. Indeed, we have
\beq
H^3 (T^4, C^*) \simeq {\rm Hom} (H^1 (T^4, C), U (1)) \simeq {\rm Hom} (\Omega, U (1)) .
\eeq
Furthermore, the pairing on $C$ induces an isomorphism $C \simeq C^*$, so $e$ can also be regarded as an element of $H^3 (T^4, C)$.

We can now describe the relationship between $(2, 0)$ theory and Yang-Mills theory in somewhat more detail. By the K\"unneth isomorphism
\beq
H^3 (T^5, C) \simeq H^2 (T^4, C) \oplus H^3 (T^4, C) ,
\eeq
the 't~Hooft flux 
\beq
f = f_{123} \lambda^1 \cup \lambda^2 \cup \lambda^3 + \ldots + f_{345} \lambda^3 \cup \lambda^4 \cup \lambda^5 \in H^3 (T^5, C)
\eeq
of the $(2, 0)$ theory decomposes into the magnetic and electric 't~Hooft fluxes $m$ and $e$ of the Yang-Mills theory:
\beq
f = m \cup \lambda^5 + e .
\eeq
Similarly, under the K\"unneth isomorphism
\beq
H^1 (T^5, \R)  \simeq H^0 (T^4, \R) \oplus H^1 (T^4, \R) ,
\eeq
the five-dimensional  momentum $p$ decomposes into the instanton number $k$ over $T^4$ and the four-dimensional  momentum $\tilde{p}$:
\beq \label{p-decomposition}
p = k \cup \lambda^5 + \tilde{p} .
\eeq
We are now in a position to understand the shifted quantization law (\ref{p-quantization}). In analogy with (\ref{mdotm}), we define a product $m \cdot e \in H^1 (T^4, \R / \Z)$ as
\beq
m \cdot e = (m \cdot e)_1 \lambda^1 + \ldots + (m \cdot e)_4 \lambda^4  , 
\eeq
where e.g.
\beq
(m \cdot e)_1 = m_{12} e_{134} + m_{13} e_{142} + m_{14} e_{123} \in \R / \Z
\eeq
and similarly for the components $(m \cdot e)_2, \ldots, (m \cdot e)_4$. Consider now a continuous spatial translation along a closed curve representing some homology class
\beq
\tilde{\lambda} \in H_1 (T^4, \Z)  \simeq \tilde{\Lambda} . 
\eeq
For a bundle with magnetic 't~Hooft flux $m$, this is equivalent to a large gauge transformation parametrized by
\beq \label{omega-class}
\omega = m [\tilde{\lambda}] \in H^1 (T^4, C) \simeq \Omega , 
\eeq
i.e. the $2$-cohomology class $m$ is partially evaluated on the one-cycle $\tilde{\lambda}$ resulting in a $1$-cohomology class $\omega$. A state with electric 't~Hooft flux $e$ then transforms with a phase
\beq
\omega \cup e \in H^4 (T^4, U (1)) \simeq U (1) ,
\eeq
which in fact equals 
\beq \label{phase}
(m \cdot e) [\tilde{\lambda}] \in \R / \Z  \simeq U (1) .
\eeq
So the four-dimensional momentum $\tilde{p} \in H^1 (T^4, \R)$ obeys the quantization law
\beq
\tilde{p} - m \cdot e \in H^1 (T^4, \Z)  \simeq \tilde{\Lambda}^* .
\eeq
Together with the relation (\ref{fractional_instanton_number}), this is equivalent to (\ref{p-quantization}), where the components of 
\beq
f \cdot f = (f \cdot f)_1 \lambda^1 + \ldots + (f \cdot f)_5 \lambda^5 \in H^1 (T^5, \R / \Z)
\eeq
are given by
\beq
(f \cdot f)_1 = f_{123} f_{145} + f_{124} f_{153} + f_{125} f_{134} \in \R / \Z
\eeq
and similarly for $(f \cdot f)_2, \ldots, (f \cdot f)_5$. Again, although we have expressed various quantities relative to the chosen basis $\lambda_1, \ldots, \lambda_5$ of $\Lambda$, the formalism is actually covariant under the $\SL_5 (\Z)$ mapping class group of $T^5$.

\subsection{The fields and the action}
For a given gauge bundle $P$, we introduce an associated vector bundle over $T^4$
\beq
\ad (P) = P \times_{\ad} \gadj ,
\eeq
where $\ad$ denotes the adjoint action of $\Gadj$ on its Lie algebra $\gadj$ (which of course equals the Lie algebra of $G$). Maximally supersymmetric Yang-Mills theory on $\R \times T^4$ with gauge group $\Gadj$ contains the following fundamental fields: 
\begin{itemize}
\item
A connection $D$ on $P$, locally represented by a connection one-form $A$ with values in $\ad (P)$. (We work in temporal gauge, so the time-component of the connection one-form is identically zero.) The magnetic field strength $F = d A + A \wedge A$ is a global section of $\Omega^2 (T^4) \otimes \ad (P)$. 
\item
Five sections $\phi$ of $\ad (P)$ transforming in the ${\bf 5}$ representation of $\Sp (4)$ .
\item
Four fermionic sections $\psi$ of $S \otimes \ad (P)$, where $S = S^+ \oplus S^-$ is the sum of the positive and negative chirality spinor bundles over $T^4$ (for the trivial spin structure). They transform in the ${\bf 4}$ representation of $\Sp (4)$.
\end{itemize} 

The complete Lagrangian is most easily obtained by dimensional reduction from $1+9$ to $1+4$ dimensions \cite{Brink-Schwarz-Scherk}. But to begin with, we will focus our attention on the terms involving only the connection one-form $A$, i.e.
\beq
\frac{1}{2 g^2} \int_{T^4} \Tr \left(\dot{A} \wedge * \dot{A} - F \wedge * F  \right) ,
\eeq
where $\Tr$ denotes a suitably normalized bilinear form on $\gadj$, and 
\beq
* : \; \Omega^k (T^4) \rightarrow \Omega^{4 - k} (T^4)
\eeq
is the Hodge duality operator constructed from the flat metric on $T^4$. (Time derivatives are denoted with a dot.) The coupling constant $g$ is related to the circumference of the $S^1$ factor of $T^5 = T^4 \times S^1$, i.e. to the magnitude of the lattice vector $\lambda_5$:
\beq
g^2 = | \lambda_5 | .
\eeq
Here we have for simplicity assumed that $\lambda_5$ is orthogonal to the subspace $\R^4$ spanned by the rank four lattice $\tilde{\Lambda}$. Otherwise, the Lagrangian would contain a further CP-violating term, analogous to the familiar theta-angle term in $(1+3)$-dimensional Yang-Mills theory.

The canonical conjugate to the connection one-form $A$ is a section $E$ (the electric field strength) of $\Omega^1 \otimes \ad (P)$ given by
\beq
E = \frac{1}{g^2} \dot{A} .
\eeq
The Hamiltonian becomes
\beq
H = \frac{g^2}{2} \int_{T^4}  \Tr \left( E \wedge * E \right) + \frac{1}{2 g^2} \int_{T^4} \Tr \left( F \wedge * F \right) ,
\eeq
and the momentum corresponding to a translation by a constant spatial vector $v$ on $T^4$ is
\beq
\iota_v \tilde{p} = \int_{T^4} \Tr \left( \iota_v F \wedge * E \right) .
\eeq

One way to view this theory is to regard it as describing a fictitious particle of mass $\mu = \frac{1}{g^2}$ moving on the flat infinite-dimensional Euclidean space of connections $A$, under the influence of a potential
\beq \label{potential}
V = \frac{1}{2 g^2} \int_{T^4} \Tr (F \wedge * F) .
\eeq

\setcounter{equation}{0}
\section{The weak coupling limit}
\subsection{The energy bound}
For a fixed non-zero value of the instanton number
\beq
k = \frac{1}{2} \int_{T^4} \Tr (F \wedge F) ,
\eeq
the potential (\ref{potential}) is bounded from below by
\beq
V \geq \frac{| k |}{g^2} .
\eeq
For positive (negative) $k$, the bound is saturated for connections $A$ with anti self-dual (self-dual) curvature $F$, i.e. $F = - * F$ ($F = *F$). Furthermore, by the Cauchy-Schwarz inequality, the magnitude square $| \tilde{p} |^2$ of the four-dimensional momentum is then bounded from above by
\beq
| \tilde{p} |^2 = \sum_v (\iota_v \tilde{p})^2 \leq \int_{T_4} \Tr (F \wedge * F) \int_{T^4} \Tr \left( E \wedge * E \right) = | k | \int_{T^4} \Tr \left( E \wedge * E \right) ,
\eeq
where $v$ is summed over an orthonormal basis of tangentvectors to $T^4$. So for fixed positive $k$ and fixed $\tilde{p}$, the energy is bounded from below by
\beq \label{energy-bound}
H \geq  \frac{k}{g^2} + \frac{g^2}{2 k} | \tilde{p} |^2  + \cO (g^4) .
\eeq
This agrees with the magnitude $| p |$ of the five-dimensional momentum $p$ decomposed as in (\ref{p-decomposition}) into components $\tilde{p}$ and $k$ along $T^4$ and $S^1$ respectively, and thus confirms the correctness of this decomposition. 

When $k = 0$, we instead have the inequality
\bea
H & \geq & \int_{T^4} \Tr \left( \frac{1}{2 g^2} \iota_v F \wedge * \iota_v F + \frac{g^2}{2} E \wedge * E \right) \cr
& = & \iota_v \tilde{p} + \frac{1}{2} \int_{T^4} \Tr \left( \frac{1}{g} \iota_v F - g E \right) \wedge * \left( \frac{1}{g} \iota_v F - g E \right) \cr
& \geq & \iota_v \tilde{p} ,
\eea
where $v$ is an arbitrary tangent vector on $T^4$. So the energy bound in this case is
\beq
H \geq | \tilde{p} | ,
\eeq
again in agreement with (\ref{p-decomposition}).

\subsection{The instanton moduli space}
For given values of the characteristic classes $k$ and $m$ with $k > 0$ ($k < 0$), we let $\cM_{k, m}$ denote the corresponding moduli space of (anti) instanton solutions. In the weak coupling limit $g \rightarrow 0$, the theory formally reduces to a supersymmetric quantum mechanical model describing a particle of mass $\mu = \frac{1}{g^2}$ moving on $\cM_{k, m}$. (The zero-point fluctuations of the modes of the connection one-form $A$ transverse to $\cM_{k, m}$ together with the contributions of the scalar fields $\phi$ cancel against the contributions of the fermionic fields $\psi$, since the corresponding eigenvalues agree \cite{dAdda-diVecchia}.) But this model is not easy to analyze, one reason being that, according to (\ref{energy-bound}), the energy gap between states of different momenta $\tilde{p}$ vanishes in the weak coupling limit. 

We will thus not attempt a complete treatment of this quantum mechanics, but content ourselves with a few remarks on its degrees of freedom. In the weak coupling limit $g \rightarrow 0$, the terms in the Yang-Mills Lagrangian involving the scalar fields $\phi$ and the spinor fields $\psi$ become
\beq
\int_{T^4} \Tr \left( \dot{\phi} \wedge * \dot{\phi} - D \phi \wedge * D \phi \right) .
\eeq
and
\beq
\int_{T^4} {\rm Vol}_{T^4} \Tr \left(\bar{\psi} (\gamma^0 \dot{\psi} + \gamma \cdot D \psi)  \right) 
\eeq
respectively. Here $\gamma^0$ and $\gamma$ denote the time-like and spatial Dirac matrices. At a generic point $A$ in $\cM_{k, m}$, the scalar Laplacian
\beq
\Delta = * D * D : \; \Gamma(\ad (P)) \rightarrow \Gamma(\ad (P))
\eeq
is strictly positive,  so we need not take the scalar fields into account. For positive $k$, there are no negative chirality zero modes for the spatial Dirac operator 
\beq
\gamma \cdot D : \; \Gamma (S \otimes \ad (P)) \rightarrow \Gamma (S \otimes \ad (P)) .
\eeq
The number of positive chirality zero modes is thus given by the Atiyah-Singer index theorem as $c_2 (\ad (P))$. (The signature and Euler characteristic of $T^4$ both vanish.) 

Let now $\psi$ be a positive chirality zero-mode, i.e. a solution to the spatial Dirac equation
\beq \label{Dirac}
\gamma \cdot D \psi = 0 .
\eeq
We can then construct two tangent vectors $\delta A$ to $\cM_{k, m}$ at the point $A$. In background Coulomb gauge $D \cdot \delta A = 0$, they take the form
\beq
\delta A = \bar{\eta} \gamma \psi ,
\eeq
where $\bar{\eta}$ is an arbitrary constant spinor of positive chirality. Indeed, the induced variation $\delta F$ of the field strength
\beq
\delta F = \bar{\eta} ( \gamma \wedge D) \psi
\eeq
is easily seen to be anti-self dual by using the chirality condition on $\bar{\eta}$ and the Dirac equation (\ref{Dirac}). In fact, $\cM_{k, m}$ is a curved hyperk\"ahler manifold of real dimension
\beq
\dim \cM = p_1(\ad (P)) = 2 c_1 (\ad (P)) ,
\eeq
so, since there are two linearly independent choices of $\bar{\eta}$, these $\delta A$ span the whole tangent space of $\cM_{k, m}$ at $A$. 

We may construct two distinguished fermionic zero modes $\psi$ as the Clifford product of $F$ and an arbitrary constant spinor $\kappa$ of positive chirality:
\beq
\psi = F \kappa .
\eeq
Indeed, it follows from the Bianchi identity $D F = 0$ and the anti self-duality of $F$ that such a $\psi$ fulfills (\ref{Dirac}). The corresponding tangent vectors $\delta A$ to $\cM_{k, m}$ at $A$ are 
\beq
\delta A = \iota_v F ,
\eeq
where the constant vector $v$ on $T^4$ is given by
\beq
v = \bar{\eta} \gamma \kappa .
\eeq
Using the Bianchi identity $D F = 0$, the induced variation of the field strength comes out to be given by the Lie derivative of $F$ along $v$:
\beq
\delta F = \cL_v F ,
\eeq
so this tangent vector $\delta A$ represents a translation of the instanton configuration on $T^4$. Such translations generically act non-trivially on $\cM_{k, m}$, which thus can be seen as a fibration with $T^4$ fiber over a hyper K\"ahler manifold $\cM^\prime_{k, m}$ of real dimension $4 (n k - 1)$. (It should be noted, though, that $\cM_{k, m}$ might be empty \cite{Braam-vanBaal}.) The wave function of a BPS state of four-dimensional momentum $\tilde{p}$ is (locally) constant on $\cM^\prime_{k, m}$ and depends on the fiber coordinates $\tilde{x} \in T^4 = \R^4 / \tilde{\Lambda}$ as $\exp (\tilde{p} \cdot \tilde{x})$. Because of the quantization of the fermionic zero-modes, this will in fact be a multi-component wave function. 

\setcounter{equation}{0}
\section{Flat connections}
A bundle $P$ of vanishing instanton number, i.e. 
\beq
k = 0 ,
\eeq
necessarily has a magnetic 't~Hooft flux $m \in H^2 (T^4, C)$ obeying
\beq
m \cdot m = 0 . 
\eeq
For such a bundle, the potential (\ref{potential}) attains its minimum value $V = 0$ for connections $A$ with vanishing field strength
\beq
F = 0 .
\eeq
In the weak coupling limit $g \rightarrow 0$, the theory thus localizes on a neighborhood of the moduli space $\cM_{0, m}$ of such flat connections. 

The moduli spaces $\cM_{0, m}$ are described in \cite{Henningson} for the $A$- and $D$-series. (For a more general theoretical discussion, which however focuses on bundles over $T^3$ rather than $T^4$, see \cite{Borel-Friedmann-Morgan}.) In general, $\cM_{0, m}$ is a disconnected sum of several components:
\beq
\cM_{0, m} = \bigcup_\alpha \cM_\alpha .
\eeq
(The range of the label $\alpha$ depends on the magnetic 't~Hooft flux $m$.) Each component $\cM_\alpha$ is of the form
\beq
\cM_\alpha = (T^{r_\alpha} \times T^{r_\alpha} \times T^{r_\alpha} \times T^{r_\alpha}) / W_\alpha 
\eeq
for some number $r_\alpha$ known as the rank of the component, and some discrete group $W_\alpha$, which acts on the torus $T^{r_\alpha}$. 

The simplest example of such a component is obtained for an arbitrary group $\Gadj$ by considering a topologically trivial bundle $P$, i.e.  not only $k = 0$ but also $m = 0$. There is then a component $\cM_0$ of $\cM_{0, m}$ for which $r_0$ equals the rank of $\Gadj$, the torus $T^{r_0}$ is a maximal torus of $\Gadj$, and the discrete group $W_0$ is the corresponding Weyl group. But even for such a trivial bundle $P$, there are in general also other components $\cM_\alpha$ of $\cM_{0, m}$. 

Returning now to the general case, the moduli space $\cM_{0, m}$ of flat connections may be parametrized by the holonomies
\beq
U_\tx \in {\rm Hom} (\pi_1 (T^4), \Gadj)
\eeq
of the connection $D$ based at some point $\tilde{x} \in T^4$ modulo simultaneous conjugation by elements of $\Gadj$. (Such conjugations represent the connected component $\Omega_0$ of the group $\tilde{\Omega}$ of gauge transformations.) A concise way of describing the holonomy $U_\tx$ is to evaluate it on the basis elements $\lambda_i$, $i = 1, \ldots, 4$ of 
\beq
\tilde{\Lambda} \simeq H_1 (T^4, \Z) \simeq \pi_1 (T^4) .
\eeq
The resulting group elements $U_\tx [\lambda_i]$ commute in $\Gadj$. But an arbitrary lifting $\hat{U}_\tx [\lambda_i]$ of them to the simply connected group $G$ is in general only almost commuting in the sense that
\beq \label{ACR}
\hat{U}_\tx [\lambda_i] \hat{U}_\tx [\lambda_j] (\hat{U}_\tx [\lambda_i])^{-1} (\hat{U}_\tx [\lambda_j])^{-1} = m_{ij} ,
\eeq
where $m_{ij} \in C$ denotes the corresponding component in the expansion of the magnetic 't~Hooft flux $m$. 

Letting $\tilde{x}$ vary over $T^4$ and evaluating the holonomy $U_\tx$ on a fixed cycle $\tilde{\lambda} \in \tilde{\Lambda}$ gives a covariantly constant section $U_\tx [\tilde{\lambda}]$ of $\Ad (P)$, i.e. 
\beq
D U_\tx [\tilde{\lambda}] \equiv d U_\tx [\tilde{\lambda}] + U_\tx [\tilde{\lambda}] A - A U_\tx [\tilde{\lambda}] = 0 .
\eeq
In other words, a large gauge transformation parametrized by $U_\tx [\tilde{\lambda}]$ leaves the connection one-form $A$ invariant:
\beq
A \rightarrow U_\tx [\tilde{\lambda}] A (U_\tx [\tilde{\lambda}])^{-1} + d U_\tx [\tilde{\lambda}] (U_\tx [\tilde{\lambda}])^{-1} = A .
\eeq
Finally, we note that the topological class of this gauge transformation is given by (\ref{omega-class}).

\subsection{Components of positive rank}
The quantization of the theory on a component $\cM_\alpha$ of positive rank $r_\alpha > 0$, is rather subtle. At a general point on this component, the holonomies $U$ spontaneously break the gauge group $\Gadj$ to a subgroup of rank $r_\alpha$. Generically, the Lie algebra $h$ of this unbroken subgroup is abelian, but in general it may be of the form
\beq
h \simeq s \oplus u (1)^r ,
\eeq
for some number $r$, $0 \leq r \leq r_\alpha$, and some semi-simple algebra $s$ of rank $r_\alpha - r$. Given such an algebra $h$, we let $\cM^h$ denote the closure of the corresponding subspace of $\cM_{0, m}$. In a neighborhood of $\cM^h$, the degrees of freedom corresponding to $s$ are modeled by supersymmetric quantum mechanics with 16 supercharges based on the Lie algebra $s$ \cite{Witten00}. The latter theory has no mass-gap, but is believed to have a finite dimensional linear space $V_s$ of normalizable zero-energy states \cite{Kac-Smilga}\cite{Henningson-Wyllard}. An explicit construction of these quantum mechanical states is notoriously difficult, essentially because the system has a potential with flat valleys extending out to infinity in field space. But this property of the supersymmetric quantum mechanics implies that the Yang-Mills theory has $\dim V_s$ normalizable zero-energy states supported near $\cM^h$. The validity of this picture was confirmed in rather much detail in \cite{Henningson}.

Somehow, there should also be a spectrum of BPS states with non-zero momentum $\tilde{p}$ supported near $\cM^h$, but it is not clear to the present author precisely how this could be determined. A better understanding of this issue would certainly be very useful, and we hope that further progress can be made, but we will not pursue it here.

\subsection{Isolated flat connections}
The situation is better for an isolated flat connection $\cD$, i.e. a component $\cM_\alpha$ of $\cM$ of rank $r_\alpha = 0$. As we will now explain, fluctuations around $\cD$ can be reliably analyzed by semi-classical methods in the weak coupling limit. We thus expand the connection one-form $A$ around the connection one-form $\cA$ of $\cD$ as
\beq \label{fluctuation}
A = \cA + g a ,
\eeq
where the fluctuation $a$ is a global section of $\Omega^1 (T^4) \otimes \ad (P)$. 

We let $\Gamma (\ad (P))$ denote the space of $L^2$-sections of the vector bundle $\ad (P)$ with respect to the sesqui-linear inner product
\beq
(\alpha, \beta) = \int_{T^4} \Tr (\bar{\alpha} \wedge * \beta) .
\eeq
Our first task is to define a convenient basis of this space. By flatness of $\cD$, the covariant derivatives
\beq
i \cD_v : \Gamma (\ad (P)) \rightarrow   \Gamma (\ad (P))
\eeq
commute with each other other for different constant vector fields $v \in H_1 (T^4, \R)$ on $T^4$. We can then introduce an orthonormal basis $b_\tp$ of $\Gamma (\ad (P))$ of their simultaneous eigensections. The label $\tp$ can be thought of as taking its values in a subset $\tilde{P}$ of $H^1 (T^4, \R)$, and is defined so that the corresponding eigenvalues are $2 \pi \tp [v]$:
\beq
i \cD_v b_\tp= 2 \pi \tp [v] b_\tp 
\eeq
(We are suppressing any further labels that are possibly needed to distinguish different sections with the same eigenvalues. But for the $A$-series, which we will treat in more detail in the next section, there is in fact no such degeneracy.)  The complex conjugate of $b_\tp$ is $(b_\tp)^* = b_{-\tp}$. The $b_\tp$ are also eigensections of the adjoint action of the holonomy $U_\tx$ evaluated on a cycle $\tilde{\lambda} \in H_1 (T^4, \Z) \simeq \tilde{\Lambda}$ based at some point $\tilde{x} \in T^4$:
\beq \label{holonomy-conjugation}
U_\tx [\lambda] b_\tp (U_\tx [\lambda])^{-1} = \tilde{z} [\lambda] b_\tp ,
\eeq
where the eigenvalue $\tilde{z} [\lambda]$ is a phase determined by
\beq
\tilde{z} = \exp (2 \pi i \tp) \in H^1 (T^4, U (1)) .
\eeq
The set $\tilde{P}$ of possible values of $\tp$ is actually best described by giving the subset $\tilde{Z} \subset H^1 (T^4, U(1))$ of possible values of $\tilde{z}$. The cardinality of this set equals the dimension of the group $G$, and, since the flat connection $\cD$ is isolated, it does not comprise the trivial element of $H^1 (T^4, U(1))$. We then have
\beq \label{P-set}
\tilde{P} = \{ \tp \in H^1 (T^4, \R) | \exp 2 \pi i \tp \in \tilde{Z} \} .
\eeq

The fluctuation $a$ in (\ref{fluctuation}) can now be expanded as
\beq
a = \sum_{\tp \in \tilde{P}} a_\tp b_\tp ,
\eeq
with some coefficients $a_\tp$ that are vectors on $T^4$. In background Coulomb gauge $\cD \cdot a = 0$, $a_\tp$ is constrained by the condition $\tp \cdot a_\tp = 0$, and thus takes its values in a 3-dimensional linear space transforming as ${\bf 1} \oplus {\bf 1} \oplus {\bf 1}$ under the $\Sp (4)$ $R$-symmetry. Similarly, we expand the scalar fields $\phi$ and the spinor fields $\psi$ as
\bea
\phi & = & \sum_{\tp \in \tilde{P}} \phi_\tp b_\tp \cr
\psi & = & \sum_{\tp \in \tilde{P}} \psi_\tp b_\tp .
\eea
Here the coefficients $\phi_\tp$ are a set of spatial scalars transforming in the  ${\bf 5}$ representation, and the coefficients $\psi_\tp$  are a set of spatial spinors transforming in the ${\bf 4} \oplus {\bf 4}$ representation of $\Sp (4)$. In the $g \rightarrow 0$ weak coupling limit, the surviving terms of the Lagrangian are
\bea
\frac{1}{2 g^2} \int_{T^4} \Tr (\dot{A} \wedge * \dot{A} - F \wedge * F) & = & \sum_{\tp \in \tilde{P}} (\dot{a}_\tp \cdot \dot{a}_{-\tp} + \tp \cdot \tp a_\tp a_{-\tp}) \cr
\int_{T^4} \Tr (\dot{\phi} \wedge * \dot{\phi} - D \phi \wedge * D \phi) & = & \sum_{\tp \in \tilde{P}} ( \dot{\phi}_\tp \dot{\phi}_{-\tp} + \tp \cdot \tp \phi_\tp \phi_{-\tp} ) \cr
\int_{T^4} {\rm Vol}_{T^4} \Tr (\bar{\psi} \gamma^0 \dot{\psi} + \bar{\psi} \gamma \cdot D \psi) & = & \sum_{\tp \in \tilde{P}} (\psi_\tp \gamma^0 \dot{\psi}_{-\tp} + \psi_\tp \tp \cdot \gamma \psi_{-\tp}) .
\eea
So for each $\tp \in \tilde{P}$, there is a set of bosonic harmonic oscillators $a_\tp$ and $\phi_\tp$ with temporal frequency $| \tp |$ transforming in the representation
\beq
B = {\bf 1} \oplus {\bf 1} \oplus {\bf 1} \oplus {\bf 5} ,
\eeq
and a set of fermionic harmonic oscillators $\psi_\tp$ with the same frequency transforming in the representation
\beq
F = {\bf 4} \oplus {\bf 4} .
\eeq

\subsection{The Hilbert space}
We will now quantize the fluctuations around an isolated flat connection $\cD$. Let $\left| \cA \right>$ denote the corresponding vacuum state of vanishing energy and momentum. Acting on this state with a string of bosonic and fermionic creation operators of the harmonic oscillators associated to a single value $\tp \in \tilde{P}$ builds up a (pre-) Hilbert space $\cH_\tp$. The complete Hilbert space $\cH$ is (the Hilbert space completion of) the tensor product
\beq
\cH = \bigotimes_{\tp \in \tilde{P}} \cH_\tp .
\eeq
There is a further subtlety that needs to be considered: If $\tilde{\lambda} \in H_1 (T^4, \Z)$ is such that $m [\tilde{\lambda}] = 0$, a gauge transformation parametrized by the holonomy $U_\tx [\tilde{\lambda}]$ belongs to the connected component $\Omega_0$ of the group of gauge transformations $\tilde{\Omega}$. We must then project the Hilbert space $\cH$ onto the subspace $\cH^{\rm inv}$ of states that are invariant under such transformations. According to (\ref{holonomy-conjugation}), this is a non-trivial projection if there are $\tilde{z} \in \tilde{Z}$ such that the phase $\tilde{z} [\tilde{\lambda}]$ is non-trivial. However, as we will see in the next section, this does not happen for the $A$-series.

In the $g \rightarrow 0$ weak coupling limit, a creation operator labeled by some $\tp \in \tilde{P}$ adds an amount $| \tp |$ to the energy and an amount $\tp$ to the momentum. The total energy $E$ and momentum $\tilde{p}$ of a state constructed by acting on the vacuum $\left| \cA \right>$ with a string of creation operators labeled by $\tp_1, \ldots \tp_k \in \tilde{P}$ are thus
\bea
E & = & |\tp_1| + \ldots + |\tp_k| \cr
\tilde{p} & = & \tp_1 + \ldots + \tp_k .
\eea
If $\tp_1, \ldots, \tp_k$ are all parallel vectors, the state is light-like in the sense that $E = \left| \tilde{p} \right|$. As discussed in the introduction,  this is a necessary condition  for it to be BPS. But it is not sufficient: For a non-zero value of the coupling $g$, the above expression for the momentum  $\tilde{p}$ of the state will still be correct, but the energy $E$ might be higher than the value stated above so that $E > | \tilde{p} |$. Indeed, already the known terms in the Lagrangian can be expected to generate such corrections at higher orders in perturbation theory, and there are presumably further unknown terms in the Lagrangian of arbitrarily high powers in the fields multiplied by appropriate powers of $g$, which may give further contributions. So most of these states, while light-like at tree level, can actually be expected to be non-BPS. 

To gain more information about which states actually are BPS, we need to consider the transformation properties under the $\Sp (4)$ $R$-symmetry. For a fixed $\tp \in \tilde{P}$, the states of $\cH_\tp$ with total momentum $k \tilde{p}$ for a positive integer $k$ transform in the representation $Z_k$ given by
\beq \label{Z_k}
Z_k = F^0_a \otimes B^k_s \oplus F^1_a \otimes B^{k-1}_s \oplus \ldots \oplus F^8_a \otimes B^{k-8}_s .
\eeq
Here the subscripts $a$ and $s$ denote the alternating and symmetric products of the bosonic and fermionic representations respectively. (If $k < 8$, the terms with negative powers of $B$ are absent.) The dimension of this representation is
\bea
\dim Z_k & = & \binom{8}{0} \binom{k+7}{7} + \binom{8}{1} \binom{k+6}{7} + \ldots + \binom{8}{8} \binom{k-1}{7} \cr
& = & \frac{16}{315} (132 k + 154 k^3 + 28 k^5 + k^7) .
\eea
We will give a more precise description of $Z_k$ in the next subsection, but at the moment we just note that by supersymmetry,
\beq
Z_k = (B \oplus F) \otimes W_k
\eeq
for some representation $W_k$. We decompose
\beq
W_k = (B \oplus F) \otimes R^\prime_k \oplus R_k ,
\eeq
where the representations $R^\prime_k$ and $R_k$ are chosen so that $R^\prime_k$ is as large as possible. Thus
\beq \label{Zkdecomposition}
Z_k = (B \oplus F)^2 \otimes R^\prime_k \; \oplus \; (B \oplus F) \otimes R_k ,
\eeq
The states transforming according to the second term must be BPS. Presumably most of the states transforming according to the first term are non-BPS, unless some additional symmetry or other principle that we have not taken into account forces them to be BPS. Somewhat more cautiously, our conclusion might be phrased as follows: While the BPS states are not necessarily given by the representation $R_k$ appearing in (\ref{Zkdecomposition}), at least the true BPS representation agrees with $R_k$ when interpreted as an element in the quotient ring
\beq \label{quotient-ring}
R_{\mathbb C} (\Sp (4)) / I .
\eeq
Here $R_{\mathbb C} (\Sp (4))$ is the representation ring  of $\Sp (4)$, and $I$ is the ideal generated by the representation $B \oplus F$. 

Sofar, we have considered the states of a single factor $\cH_\tp$ for $\tp \in \tilde{P}$. A state of total momentum $\tilde{p} \in H^1 (T^4, \R)$ (which is not necessarily an element of $\tilde{P}$) in the complete Hilbert space $\cH$ is clearly non-BPS if oscillators of different values $\tp_1, \ldots, \tp_k \in \tilde{P}$ are excited, since such a state is not even light-like at tree level. But if only $k$ oscillators of momentum $\tilde{p} / k \in \tilde{P}$ are excited, we can apply our previous reasoning and argue that (at least) a set of states transforming as $(B \oplus F) \otimes R_k$ are BPS. Taking all possible values of $k$ into account, we thus conjecture that the BPS states of some fixed momentum $\tilde{p} \in H^1 (T^4, \R)$ transform as $(B \oplus F) \otimes R$ where
\beq \label{Rsum}
R = \sum_{k | \tilde{p}} R_k .
\eeq
Here the sum is over all positive integers $k$ that divide $\tilde{p}$ in the sense that $\tilde{p} / k \in \tilde{P}$. Again, we have a rigorous statement in the quotient ring (\ref{quotient-ring}).
 
 \subsection{Some $\Sp (4)$ representation theory}
We will now work out the precise form of the representation $R_k$ for a given value of $k$. An irreducible representation $V_{k_1, k_2}$ of the $\Sp (4)$ $R$-symmetry group is labelled by two non-negative integers $k_1$ and $k_2$ (the Dynkin labels). By the Weyl dimension formula, the dimension of this representation is
\beq
\dim V_{k_1, k_2} = \frac{1}{6} (1 + k_1) (1 + k_2) (2 + k_1 + k_2) (3 + 2 k_1 + k_2) .
\eeq
We have e.g.
\bea
V_{0, 0} & = & {\bf 1} \cr
V_{0, 1} & = & {\bf 4} \cr
V_{1, 0} & = & {\bf 5} .
\eea
In particular, the symmetric powers of the representation ${\bf 5}$ are given by 
\beq
{\bf 5}^k_s = \left\{ 
\begin{array}{ll}
V_{0,0} \oplus V_{2,0} \oplus \ldots \oplus V_{k,0} &  {\rm for \;\; even \;\;} k \cr
V_{1,0} \oplus V_{3,0} \oplus \ldots \oplus V_{k,0} &  {\rm for \;\; odd \;\;} k .
\end{array}
\right.
\eeq
It follows that
\beq
B^k_s = \bigoplus_{k_1 = 0}^k N_{k - k_1} \times V_{k_1, 0} ,
\eeq
where the multiplicities $N_l$ are given by
\beq
N_l = \left\{
\begin{array}{ll}
\frac{1}{24}  (24 + 34 l + 15 l^2 + 2 l^3) & {\rm for \;\; even \;\;} l \cr
\frac{1}{24}  (21 + 34 l + 15 l^2 + 2 l^3) & {\rm for \;\; odd \;\;} l .
\end{array}
\right.
\eeq

After some more work, it transpires that the representations $R_k$ and $R^\prime_k$ defined above are (almost) given by
\bea \label{dim-Rk}
R_k & =  & 3 \times V_{k-2, 0} \cr
& & {} \oplus V_{k-1, 0} \cr
& & {} \oplus \bigoplus_{k_1 = 0}^{k-3} (4 k - 4 k_1 - 6) \times V_{k_1, 0} \cr
& & {} \oplus \bigoplus_{k_1 = 0}^{k - 2} (2 k - 2 k_1 - 4) \times V_{k_1, 1}   
\eea
and
\beq \label{dim-Rkprime}
R^\prime_k = \bigoplus_{k_1 = 0, k-4} N_{k - 4 - k_1} \times V_{k_1, 0} .
\eeq
Their dimensions are
\beq 
\dim R_k = \frac{1}{15} (8 k + 5 k^3 + 2 k^5)
\eeq
and
\beq 
\dim R_k^\prime = \frac{1}{7!} (k-3)(k-2)(k-1)k(k+1)(k+2)(k+3) .
\eeq
These representations obey equation (\ref{Zkdecomposition}), but not quite the requirement that $R^\prime_k$ be as large as possible. In fact, for $k \geq 3$ it is possible to add up to $k - 2$ trivial representations $V_{0, 0}$ to $R^\prime_k$ and remove the same number of representations 
\beq
B \oplus F = 3 \times V_{0,0} \oplus V_{1,0} \oplus 2 \times V_{0,1}
\eeq
 from $R_k$. I do not know of any argument to prove that these terms indeed correspond to BPS states, and thus should be included in $R_k$ rather than $R_k^\prime$, except that the formulas (\ref{dim-Rk}) and (\ref{dim-Rkprime}) look simpler that way.

\setcounter{equation}{0}
\section{The $A$-series}
To be able to use the analysis of the previous section, we must consider a gauge group $\Gadj$ and a magnetic 't~Hooft flux $m \in H^2 (T^4, C)$ such that the corresponding moduli space $\cM_{0, m}$ of flat connections only consists of components $\cM_\alpha$ of rank $r_\alpha = 0$, i.e. of isolated flat connections. This will restrict us to the $A$-series and prime values of the magnetic 't~Hooft flux. But again, one could hope to eventually be able to analyze also components of positive rank, so that arbitrary groups and 't~Hooft fluxes could be treated.

Consider thus the case $\Phi = A_{n-1}$ for some positive integer $n$. The corresponding simply connected group $G = \SU (n)$ consists of unimodular $n \times n$ matrices. Its center subgroup $C$ consists of matrices of the form $\exp (2 \pi i c / n) \id_n$, where $c \in \Z_n \simeq C$. The inner product on $C$ is given by
\beq
c \cdot c^\prime = \frac{1}{n} c c^\prime \in \R / \Z ,
\eeq
for $c, c^\prime \in \Z_n$.

For $m \in H^2 (T^4, \Z_n)$, we define the $\SL_4 (\Z)$ invariant $u$ as the greatest common divisor of the components of $m$ and $n$:
\beq
u = \gcd (m_{12}, \ldots, m_{34}, n) .
\eeq
We can then write 
\beq
m = u m^\prime
\eeq
for some $m^\prime \in H^2 (T^4, \Z_{n / u})$, and define a further $\SL_4 (\Z)$ invariant $m^\prime \cdot m^\prime \in \Z_{n / u}$ as
\beq
m^\prime \cdot m^\prime = m^\prime_{12} m^\prime_{34} + m^\prime_{13} m^\prime_{42} + m^\prime_{14} m^\prime_{23}  .
\eeq
One can show (see e.g. \cite{Henningson}) that all the connected components $\cM_\alpha$ of the moduli space $\cM_{0, m}$ of flat connections have the same rank $r_\alpha$ given by
\beq
r_\alpha =  u \left/ \frac{n/u}{\gcd (m^\prime \cdot m^\prime, n/u)} \right. - 1,
\eeq
provided that this quantity is an integer. (Otherwise, $\cM_{0, m}$ is empty.) So we get $r_\alpha = 0$ if e.g. 
\beq
u = 1
\eeq
and 
\beq
m \cdot m = 0 . 
\eeq
In fact, there is then a unique isolated flat connection $\cA$. (There are, up to simultaneous conjugation, $n^2$ different quadruples $\hat{U}_\tx [\lambda_i] \in G$, $i = 1, \ldots, 4$ that fulfill the almost commutation relations (\ref{ACR}), but they project to a unique quadruple $U_\tx [\lambda_i] \in \Gadj$.) We can describe this connection via the adjoint action of its holonomies on the space of sections $\Gamma (\ad (P))$ as in (\ref{holonomy-conjugation}). The set $\tilde{Z}$ in which $\tilde{z}$ takes its values can then be regarded as a subset of the cohomology group
\beq
H^1 (T^4, \Z_n) \simeq H^1 (T^4, \frac{1}{n} \Z / \Z) \subset H^1 (T^4, \R / \Z) \simeq H^1 (T^4, U (1)) .
\eeq
In fact,
\beq
\tilde{Z} = \left\{ \tilde{z} \in H^1 (T^4, \Z_n) \Big| \tilde{z} \neq 0, \; \tilde{z} \wedge m = 0 \right\} .
\eeq
Note that the cardinality of $\tilde{Z}$ equals the dimension $n^2 - 1$ of $G = \SU (n)$. It is not difficult to check that for a given value of $\tilde{z} \in \tilde{Z}$, there exists a corresponding value of the electric 't~Hooft flux $e \in H^3 (T^4, \Z_n)$, unique modulo terms of the form $m \wedge t$ for $t \in H^1 (T^4, \Z_n)$, such that
\beq
\tilde{z} = m \cdot e \in H^1 (T^4, \frac{1}{n} \Z / \Z) .
\eeq
According to (\ref{phase}), it is then consistent to declare that the corresponding harmonic oscillators are invariant under topologically trivial gauge transformations, and transform under `large' gauge transformations according to $e$. There is thus no need to project onto states invariant under gauge transformations in the connected component $\Omega_0$. The four-dimensional momentum $\tilde{p}$ of the harmonic oscillators takes its values in the set $\tilde{P}$ defined in (\ref{P-set}):
\beq
\tilde{P} = \left\{ \tilde{p} \in H^1 (T^4, \frac{1}{n} \Z) \Big| n \tilde{p} \neq 0 , \; (n \tilde{p}) \wedge m = 0\right\} ,
\eeq
where $n \tilde{p} \in H^1 (T^4, \Z)$.

\subsection{The case of $n$ prime}
The conditions on $u$ and $m^\prime \cdot m^\prime$ are not as restrictive as they may seem: For each possible value of $m$, there are $n^4$ different values of $e \in H^3 (T^4, C)$, and the resulting values of $f = m + e$ actually give representatives of many $\SL_5 (\Z)$ orbits of $f$. E.g. for $n$ prime (so that $u = 1$), there are $n+1$ $\SL_4 (\Z)$ orbits of $m$:
\beq
\begin{array}{ll}
{\rm orbit} & {\rm cardinality} \cr
\hline
m = 0 & 1 \cr
m \neq 0, m \cdot m  = 0 & n^5 + n^3 - n^2 - 1 \cr
m \cdot m = 1 / n & n^5 - n^2 \cr
\ldots & \ldots \cr
m \cdot m = (n-1) / n & n^5 - n^2 \cr
\hline
& n^6
\end{array}
\eeq
But there are only 3 $\SL_5 (\Z)$ orbits of $f$:
\beq
\begin{array}{ll}
{\rm orbit} & {\rm cardinality} \cr
\hline
f = 0 & 1 \cr
f \neq 0, f \cdot f = 0 & n^7 + n^5 - n^2 - 1 \cr
f \cdot f \neq 0 & n^{10} - n^7 - n^5 + n^2 \cr
\hline
& n^{10}
\end{array}
\eeq
The relationship between these orbits is
\beq
\begin{array}{cccccc}
& m = 0 & m \cdot m = 0 & m \cdot m = 1/n & \ldots & m \cdot m = (n - 1) / n \cr
\hline
f = 0 & 1 & &  & \ldots & \cr
f \cdot f = 0 & n^4 - 1 & n^2 & & \ldots & \cr
f \cdot f \neq 0 & & n^4 - n^2 & n^4 & \ldots 
\end{array}
\eeq
where the entries denote the number of $e$-values for a given $m$ in the corresponding $\SL_4 (\Z)$ orbit that gives an $f$ in the corresponding $\SL_5 (\Z)$ orbit. So although only the $m \neq 0$, $m \cdot m = 0$ orbit has an isolated flat connection, this makes calculations possible for all $f$, except the single value $f = 0$.

Finally, we consider the action of $\SL_5 (\Z)$ on the set of pairs $(f, p)$ obeying (\ref{p-quantization}): For the $f = 0$ orbit, we have $p \in H^1 (T^5, \Z)$. There is then one orbit for each positive integer value of 
\beq
\gcd (p) = \gcd (p_1, \ldots, p_5) . 
\eeq
But as discussed above, we have no results for the corresponding BPS spectrum. Also for the $f \neq 0$, $f \cdot f = 0$ orbit, we have $p \in H^1 (T^5, \Z)$. But here there are two orbits for each positive integer value of $\gcd (p)$, distinguished by 
\beq
f \wedge (p / \gcd (p)) \in H^4 (T^5, C)
\eeq
being zero or non-zero. In both cases, the BPS spectrum could be determined as described above by choosing a decomposition (\ref{Lambda-decomposition}) of $\Lambda$ such that the component $k$ in the decomposition (\ref{p-decomposition}) of $p$ vanishes. In fact the BPS spectrum is empty when $f \wedge (p / \gcd (p)) \neq 0$. For the $f \cdot f  \neq 0$ orbit, we consider $n p \in H^1 (T^5, \Z)$ instead of $p \in H^1 (T^5, \frac{1}{n} \Z)$. There is then one orbit for each finite value of $\gcd (n p) = \gcd (n p_1, \ldots, n p_5)$. (Necessarily $f \wedge  (n p / \gcd (n p)) = 0$ in this case.) Also here, our methods allow for a determination of the BPS spectrum.

\vspace*{5mm}
This research was supported by grants from the G\"oran Gustafsson foundation and the Swedish Research Council.

\end{document}